\documentclass[fleqn,usenatbib]{mnras}

\usepackage{newtxtext,newtxmath}

\usepackage[T1]{fontenc}

\DeclareRobustCommand{\VAN}[3]{#2}
\let\VANthebibliography\thebibliography
\def\thebibliography{\DeclareRobustCommand{\VAN}[3]{##3}\VANthebibliography}

\usepackage{orcidlink}
\usepackage{graphicx}	
\usepackage{amsmath}	
\usepackage{amsfonts}
\usepackage{bm}
\usepackage{multirow} 
\usepackage{ulem}
\usepackage{comment}

\newcommand{\hmpc}{\,h^{-1}\,\mathrm{Mpc}}
\newcommand{\hgpc}{\,h^{-1}\,\mathrm{Gpc}}
\newcommand{\ihmpc}{\,h\,\mathrm{Mpc}^{-1}}
\newcommand{\lsub}{L_\mathrm{sub}}
\newcommand{\lcell}{L_\mathrm{cell}}
\newcommand{\ngrid}{N_\mathrm{grid}}
\newcommand{\nrebin}{N_\mathrm{rebin}}
\newcommand{\nbatch}{n_\mathrm{batch}}
\newcommand{\dini}{\delta_\mathrm{ini}(z=10)}
\newcommand{\dfin}{\delta_\mathrm{fin}(z=0)}
\newcommand{\drec}{\delta_\mathrm{rec}}

\newcommand{\revise}[1]{\textcolor{black}{#1}}
\newcommand{\ajn}[1]{\textcolor{magenta}{[AJN:#1]}}

\title[Optimal scales for density reconstruction]{Searching optimal scales for reconstructing cosmological initial conditions using convolutional neural networks}

\author[K. Nakashima et al.]{
Koichiro Nakashima$^{1}$\orcidlink{0009-0004-7367-3772}\thanks{E-mail: nakashima.koichiro.v2@s.mail.nagoya-u.ac.jp},
Kiyotomo Ichiki$^{1,3,4}$\orcidlink{0000-0003-1365-8568},
Atsushi J. Nishizawa$^{2,3,4}$\orcidlink{0000-0002-6109-2397}\thanks{atsushi.nishizawa@iar.nagoya-u.ac.jp}, 
Kenji Hasegawa$^{5}$
\\
$^{1} $Department of Physics, Nagoya University, Furocho, Chikusa, Nagoya, Aichi 464-8602, Japan,\\
$^2$ DX Center, Gifu Shotoku Gakuen University, Takakuwanishi, Yanaizu, Gifu, 501-6194, Japan\\
$^3$ Institute for Advanced Research, Nagoya University, Furocho, Chikusa, Nagoya, Aichi, 464-8602, Japan\\
$^4$ Kobayashi Maskawa Institute, Nagoya University, Furocho, Chikusa, Nagoya, Aichi, 464-8602, Japan\\
$^5$ National Institute of Technology, Suzuka College, Shirokocho, Suzuka, Mie 510-0294, Japan \\
}

\date{Accepted XXX. Received YYY; in original form ZZZ}

\pubyear{\the\year{}}

\begin{document}
\label{firstpage}
\pagerange{\pageref{firstpage}--\pageref{lastpage}}
\maketitle

\begin{abstract}
Reconstructing the initial density field of the Universe from the late-time matter distribution is a non-trivial task with implications for understanding structure formation in cosmology, offering insights into early Universe conditions. Convolutional neural networks (CNNs) have shown promise in tackling this problem by learning the complex mapping from non-linear evolved fields back to initial conditions. Here we investigate the effect of varying input sub-box size in single-input CNNs. We find that intermediate scales ($\lsub \sim 152\hmpc$) strike the best balance between capturing local detail and global context, yielding the lowest validation loss and most accurate recovery across multiple statistical metrics. We then propose a dual-input model that combines two sub-boxes of different sizes from the same simulation volume. 
This model significantly improves reconstruction performance, especially on small scales over the best single-input case, despite utilizing the same parent simulation box. This demonstrates the advantage of explicitly incorporating multi-scale context into the network. Our results highlight the importance of input scale and network design in reconstruction tasks. The dual-input approach represents a simple yet powerful enhancement that leverages fixed input information more efficiently, paving the way for more accurate cosmological inference from large-scale structure surveys. 
\end{abstract}




\section{Introduction}

The large-scale structure of the Universe observed today has evolved over billions of years from nearly uniform initial conditions, shaped predominantly by gravitational instability. This evolution encodes key information about fundamental components of the Universe—dark matter, dark energy, and baryonic matter—and their interactions. Recovering the initial matter distribution from its present-day state is a longstanding goal in cosmology, as it offers a clearer view of the early Universe and a more direct probe of fundamental physics.

If we could undo the gravitational evolution that shaped the current matter distribution, we would gain access to the Universe's initial conditions at the time of recombination, before structure formation distorted the original signals. One important feature from that era is the peak in the matter two-point correlation function arising from baryon acoustic oscillations (BAO), which are remnants of sound waves that propagated through the photon-baryon fluid in the early Universe and became imprinted in the matter field at recombination. This BAO scale serves as a standard ruler to trace the expansion of the Universe and constrain dark energy models \citep{Weinberg2013}. Although the BAO signature remains detectable in the late-time matter distribution, it is altered by the movement of galaxies due to gravitational clustering. Reconstructing the primordial density field would improve the accuracy of sound horizon measurements. Moreover, the early matter distribution provides a valuable window into primordial non-Gaussianity \citep{Bartolo2004} and may reveal previously unknown aspects of the Universe’s composition and governing physics.

An important motivation for reconstructing the initial density field is the possibility of performing direct comparisons of the matter distribution across different cosmic times, particularly at matched phases. If the late-time non-linear density field can be accurately “pulled back” to its linear regime, it opens the door to powerful consistency tests of structure formation. One compelling application is the Kamionkowski-Loeb method \citep{Kamionkowski1997}, which proposes measuring the large-scale gravitational potential through the polarization of CMB photons scattered in galaxy clusters. The polarization angle and amplitude are sensitive to the quadrupole of CMB temperature fluctuations as seen from each cluster \citep{Seto2000}, allowing us to reconstruct the three-dimensional density fluctuations of the universe at the recombination epoch \citep[e.g.][]{Portsmouth2004,2022PhRvD.105f3507I}.

If we can reconstruct the initial density field from present-day observations, we can directly compare it with the linear-density estimates derived at earlier epochs using this method, enabling novel cross-redshift validation of the $\Lambda$CDM paradigm and tests of late-time gravitational effects such as the integrated Sachs-Wolfe (ISW) effect.

Gravitational dynamics are well understood in the forward direction, allowing us to simulate the formation of structure with high precision. However, reversing this process to infer the initial conditions from late-time observations is fundamentally more difficult. Non-linear effects, such as gravitational collapse and shell crossing, erase information at small scales, while the long-range nature of gravity complicates the inversion even at larger scales. 

Several methods have been developed to tackle this inverse problem, ranging from perturbative techniques to iterative solvers and probabilistic forward models. \citet{Nusser1992} formulated equations for gravitational fluid motion under the Zel’dovich approximation \citep{Zel'dovich1970}, which allowed the matter distribution to be evolved backward in time. A major advance came with \citet{Eisenstein2007}, who showed that the BAO signal in galaxy clustering data can be enhanced by applying a reconstruction technique based on the continuity equation from linear perturbation theory. It has since been widely explored in both theory (e.g. \citet{Seo2008,Noh2009,Padmanabhan2009,Schmittfull2015}) and observations (e.g. \citet{Padmanabhan2009,Xu2013,Hinton2017}).

\revise{
Building on the success of these reconstruction techniques, a variety of alternative methods have recently been developed and applied more widely. \citet{Zhu2016} recovered the primordial density field by solving for the non-linear displacement field, demonstrating in 1D that suppressed linear modes could be restored and non-linear damping reduced. This framework was extended to 3D with a multigrid solver (\citet{Zhu2017}), enabling recovery of initial conditions to non-linear scales and enhancing BAO and redshift-space analyses. Subsequent work evaluated its utility: \citet{Pan2017} quantified large Fisher-information gains with a moving-mesh scheme; \citet{Wang2017} showed that isobaric reconstruction sharply reduced BAO smearing; \citet{Yu2017} demonstrated strong linear recovery from realistic halo densities; \citet{WangPen2019} analysed tracer-bias effects; and \citet{Zhu2018} extended the method to redshift space. \citet{Schmittfull2017} provided a detailed review and an iterative reconstruction algorithm, and \citet{Seljak2017} developed a hierarchical Bayesian method using optimization and perturbative expansions or \citet{Shi2018} solved the non-linear Lagrangian-to-Eulerian mapping and \citet{Wang2020} removed redshift-space distortions from galaxy clustering measurements.
}

In light of these challenges, recent studies have turned to machine learning-based approaches, particularly convolutional neural networks (CNNs; e.g. \citet{Krizhevsky2012}), to address the reconstruction task from a data-driven perspective. \citet{Mao2021} proposed a CNN-based method that operates directly on the final density field in sub-volumes of the Universe, training the model to infer the corresponding initial density distribution. Their network achieved promising results, accurately recovering large-scale modes and enhancing the signal-to-noise ratio of the BAO feature. A key advantage of their approach is its locality in configuration space, which reduces sensitivity to survey boundaries compared to traditional methods. Moreover, they demonstrated that a model trained under one cosmology generalizes well to others, highlighting the potential of CNN-based methods as flexible and robust tools for future galaxy surveys.

Building on this direction, \citet{Shallue2023} argued that 
\revise{a key challenge arises from the conflict between the long-range nature of gravitational interactions and the inherently local receptive fields of CNNs, which remain limited even with deep architectures. While enlarging the receptive field through larger kernels, additional layers, or strided/dilated convolutions is possible, each approach entails tradeoffs in computational cost or small-scale resolution.}
Instead, they developed a hybrid approach: first applying standard first-order reconstruction to reverse large-scale bulk flows, then using a CNN to learn the residual non-linear transformation. This two-step strategy significantly enhances reconstruction accuracy, especially at smaller scales ($k\sim0.4~h,\mathrm{Mpc}^{-1}$), and mitigates redshift-space distortions. 

In a related study, \citet{Chen2023} extended this hybrid approach by systematically investigating how the performance of CNN-based reconstruction depends on the choice of reconstruction algorithm, its parameters, and the level of shot noise. Using density fields reconstructed from the N-body simulations, they trained an eight-layer CNN to enhance the correlation with the true initial conditions, achieving accuracy up to $k \sim 0.5h~\mathrm{Mpc}^{-1}$ in configuration space. Their method also improved the removal of redshift-space distortions and demonstrated robustness across cosmological models, though its effectiveness diminished with increasing shot noise.
Moreover, \citet{Parker2025} proposed a hybrid method combining standard BAO reconstruction with a CNN model trained on partitioned subgrids to learn small-scale corrections, using mock halo and galaxy catalogs derived from N-body simulations. Notably, their method achieves robust performance in both configuration and redshift space, generalizes well to larger volumes without retraining, and yields enhanced constraints on the acoustic scale.
These works highlight how combining physical insight with deep learning architectures can overcome the limitations of purely data-driven models.

In this paper, we aim to reconstruct the initial linear matter density field directly from the non-linearly evolved dark matter field using CNNs trained on simulated data. Rather than focusing on sharpening specific features such as the baryon acoustic oscillations (BAO), our goal is a high-fidelity reconstruction of the full initial density field across a range of scales. To this end, we explore the effectiveness of deep convolutional neural networks in capturing the complex, non-linear mapping between the late-time and initial matter distributions, with particular attention to the impact of input scale and the incorporation of multi-scale information. Building on prior efforts, we investigate how the choice of spatial scale for the CNN input influences reconstruction performance.
\revise{Furthermore, we introduce a new method that enables CNNs alone, unlike the hybrid strategies of \citet{Shallue2023} and \citet{Parker2025}, to effectively incorporate multi-scale context. Specifically, alongside the conventional small input boxes that capture local structure, we provide coarser-resolution, larger boxes, allowing the network to jointly learn local information and global context.}

The structure of this paper is as follows. In Section~\ref{sec:methods}, we describe our methodology, including the dataset, input preprocessing, CNN architecture, and training procedures. Section~\ref{sec:results} presents the main results, examining how reconstruction accuracy varies with input scale and evaluating the performance of the dual-input model using multiple statistical metrics. Section~\ref{sec:discussions} offers a discussion of the implications and limitations of our approach. Finally, Section~\ref{sec:conclisions} summarizes our conclusions and outlines potential directions for future work. 

\section{Methods}
\label{sec:methods}
Our method follows the framework proposed by \citet{Mao2021}, which employs a network-based method to reconstruct cosmological initial density fields from non-linear evolved density fields. In this section, we briefly describe the structure of our network model, the preparation of the data set, and the training procedure used to optimize the model.


\begin{table}
 \centering
\begin{tabular}{lcccc}
\hline 
Layer & Kernel size & Output shape & Stride \\
\hline \hline
input & None & ($\nbatch$, 39, 39, 39, 1) & None  \\\hline
conv1 & (3, 3, 3) &($\nbatch$, 20, 20, 20, 32)& (2, 2, 2) \\ \hline
conv2 & (3, 3, 3) &($\nbatch$, 20, 20, 20, 32) & (1, 1, 1)   \\\hline
conv3 & (3, 3, 3) &($\nbatch$, 10, 10, 10, 64) & (2, 2, 2)    \\\hline
conv4 & (3, 3, 3) &($\nbatch$, 10, 10, 10, 64) & (1, 1, 1)  \\\hline
conv5 & (3, 3, 3) &($\nbatch$, 5, 5, 5, 128) & (2, 2, 2) \\\hline
conv6 & (3, 3, 3) &($\nbatch$, 5, 5, 5, 128)  & (1, 1, 1)  \\\hline
conv7 & (1, 1, 1) &($\nbatch$, 5, 5, 5, 128)  & (1, 1, 1)  \\\hline
mean  & None &($\nbatch$, 128)   & None  \\\hline
fully connected & None &($\nbatch$, 1) & None  \\\hline\hline
\end{tabular}
\caption{The architecture of the single-input convolutional neural network used in this work, same as \citet{Mao2021}. For the convolutional layers, the input and output tensors follow the format \texttt{[batch size, depth, height, width, channels]}. In contrast, for the fully connected layers, the tensor shape is \texttt{[batch size, channels]}. We vary the batch size $\nbatch$ using values of 32 (iteration=[0,1$]\times10^5$), 128 (iteration=[1,3]$\times10^5$) and 512 (iteration=[3,5]$\times10^5$) during training. To prevent excessive reduction in spatial dimensions due to convolution, padding is applied in layers \texttt{conv1} through \texttt{conv6}. All convolutional layers are followed by a ReLU activation function~\citep{Nair2010}. The output of the final convolutional layer is averaged across spatial dimensions before being passed to a fully connected layer that produces the final scalar prediction.}
\label{tab:cnn_single}
\end{table}

\subsection{Convolutional Neural network}
\label{subsec:cnn} 
A neural network (e.g. \citet{LeCun2015,Goodfellow2016}) is a parametric function composed of layers of interconnected units, or neurons, designed to approximate complex mappings between input and output data. Each neuron performs a weighted sum of its inputs, adds a bias term, and applies a non-linear activation function. For layer $i\in\mathbb{N}$, the transformation from an input vector $\bm{x}_{i-1} \in\mathbb{R}^n$ to an output vector 
$\bm{x}_{i} \in\mathbb{R}^m$
can be written as:
\begin{align}
    \bm{x}_i=f(\bm{W}_i\bm{x}_{i-1}+\bm{b}_i),
\end{align}
where $\bm{W}_i \in\mathbb{R}^{m\times n}$ is a weight matrix, $\bm{b}_i\in\mathbb{R}^m $ is a bias factor, and $f$ is a non-linear activation function such as the ReLU (Rectified Linear Unit;~\citet{Nair2010}). By stacking multiple layers, neural networks can learn hierarchical representations of the input data. Training a neural network consists of optimizing the parameters $\bm{W}$ and $\bm{b}$ so that the model accurately maps inputs to their corresponding outputs, given a set of input–output pairs.

Convolutional neural networks (CNNs; e.g. \citet{Krizhevsky2012}) extend standard neural networks, in which the output of the $l$-th kernel in layer $i$ is
\begin{align}
    \bm{x}^l_i=f\left(\sum_k{\bm{W}^{l,k}_i\otimes\bm{x}^k_{i-1}+\bm{b}_i^l}\right),
\end{align}
where $\otimes$ denotes the convolutional operation, the convolutional kernels $\bm{W}^{l,k}_i$ and bias factor $\bm{b}_i^l$ are the trainable parameters. In general, $\bm{x}^l_i$ is a $d$-dimensional grid.  CNNs replace the global matrix–vector operation of a standard neural network with a spatially localized sum of weighted inputs, capturing local spatial features efficiently. This structure is particularly well suited for data with inherent spatial correlations, such as cosmological density fields. In addition to convolutional layers, pooling layers are often employed to progressively reduce the spatial dimensions of feature maps. By aggregating information over local neighborhoods, typically through operations such as maximum or average pooling, these layers enhance translational invariance, suppress small-scale noise, and reduce computational complexity.


Following \citet{Mao2021}, the network inputs a 3-D density field at present and outputs the initial condition. Since the gravitational evolution is non-linear, we have to incorporate the non-local information. Therefore, we consider $39\times 39\times 39$ voxels around the target position as an input array. We denote this non-local scale as $L_{\rm sub}=39\times L_{\rm cell}$.
The size of the kernel $\bm{W}^{l,k}_i$ is fixed at $3\times3\times3$ voxels for all convolutional layers, except for the final layer, which uses a $1\times1\times1$ kernel. Instead of employing explicit pooling layers, spatial downsampling is achieved via convolutions with a stride of 2 voxels in each spatial dimension. The model consists of seven convolutional layers, each followed by a ReLU activation function~\citep{Nair2010}. The output of the final convolutional layer is averaged over the spatial dimensions and passed to a fully connected layer that produces the final scalar prediction. The architecture of the single-input network is summarised in Table~\ref{tab:cnn_single}.

In \citet{Mao2021}, the non-local scale $L_{\rm sub}=76 \hmpc$ but we extend this larger or smaller scales to find the optimal scales to reconstruct the initial density field. Different combinations of $L_{\rm cell}$ and $L_{\rm sub}$ are summarized in Table~\ref{tab:cnn_scales}. Details on how we generate the dataset for each sub-box configuration are provided in Section~\ref{subsec:dataset}.

In addition to the 
input which has single $L_{\rm sub}$,
we also implement a dual-input network in which two 
different $\lsub$'s are
used as inputs. The architecture of the dual-input network is summarised in Table~\ref{tab:cnn_dual}. Each input branch independently processes a $39^3$ voxel grid through seven convolutional layers with ReLU activations, with padding applied in the first six layers to preserve spatial dimensions. The extracted feature maps from both branches are then concatenated and passed through a series of fully connected layers to produce the final scalar prediction.
\begin{table}
\centering
\resizebox{0.48\textwidth}{!}{
\begin{tabular}{ccccc}
\hline 
\multicolumn{2}{c|}{Input density ($z=0$)} && \multicolumn{2}{c}{Target density ($z=10$)}\\ \hline
$\lsub~[\hmpc]$ & $\lcell^\mathrm{sub}~[\hmpc]$ &$\nrebin$& $\lcell^\mathrm{box}~[\hmpc]$ & $\ngrid$ \\\hline \hline
38 & 0.97 &1& 0.97 & $1024^3$ \\\hline
76 & 1.95 &1& 1.95 & $512^3$ \\\hline
114 & 2.92 &3& 0.97 & $1024^3$ \\\hline
152 & 3.90 &1& 3.90 & $256^3$ \\\hline
190 & 4.88 &5& 0.97 & $1024^3$ \\\hline
228 & 5.85 &3& 1.95 & $512^3$ \\\hline
266 & 6.83 &7& 0.97 & $1024^3$ \\\hline
304 & 7.81 &1& 7.81 & $128^3$ \\\hline
342 & 8.78 &9& 0.97 & $1024^3$ \\\hline
380 & 9.76 &5& 1.95 & $512^3$ \\\hline\hline
\end{tabular}
}
\caption{List of sub-box sizes $\lsub$, cell sizes $\lcell^{\{\mathrm{sub},\,\mathrm{box}\}}$, the size of "rebinnig" $\nrebin$ and grid numbers $\ngrid$ used for the input and target density fields. As described in Section~\ref{subsec:dataset}, we generated both input and target density fields within a 1$\hgpc$ simulation box using different grid resolutions $\ngrid$, resulting in varying cell sizes $\lcell^\mathrm{box}$. For the input field, we vary the sub-box size $\lcell^\mathrm{sub}$ by averaging over a block of $(\nrebin)^3$ cells, fixing the number of grid points within each sub-box to $39^3$.
}
 \label{tab:cnn_scales}
\end{table}
\begin{figure*}
\centering
    \centering
    \includegraphics[width=0.9\linewidth]{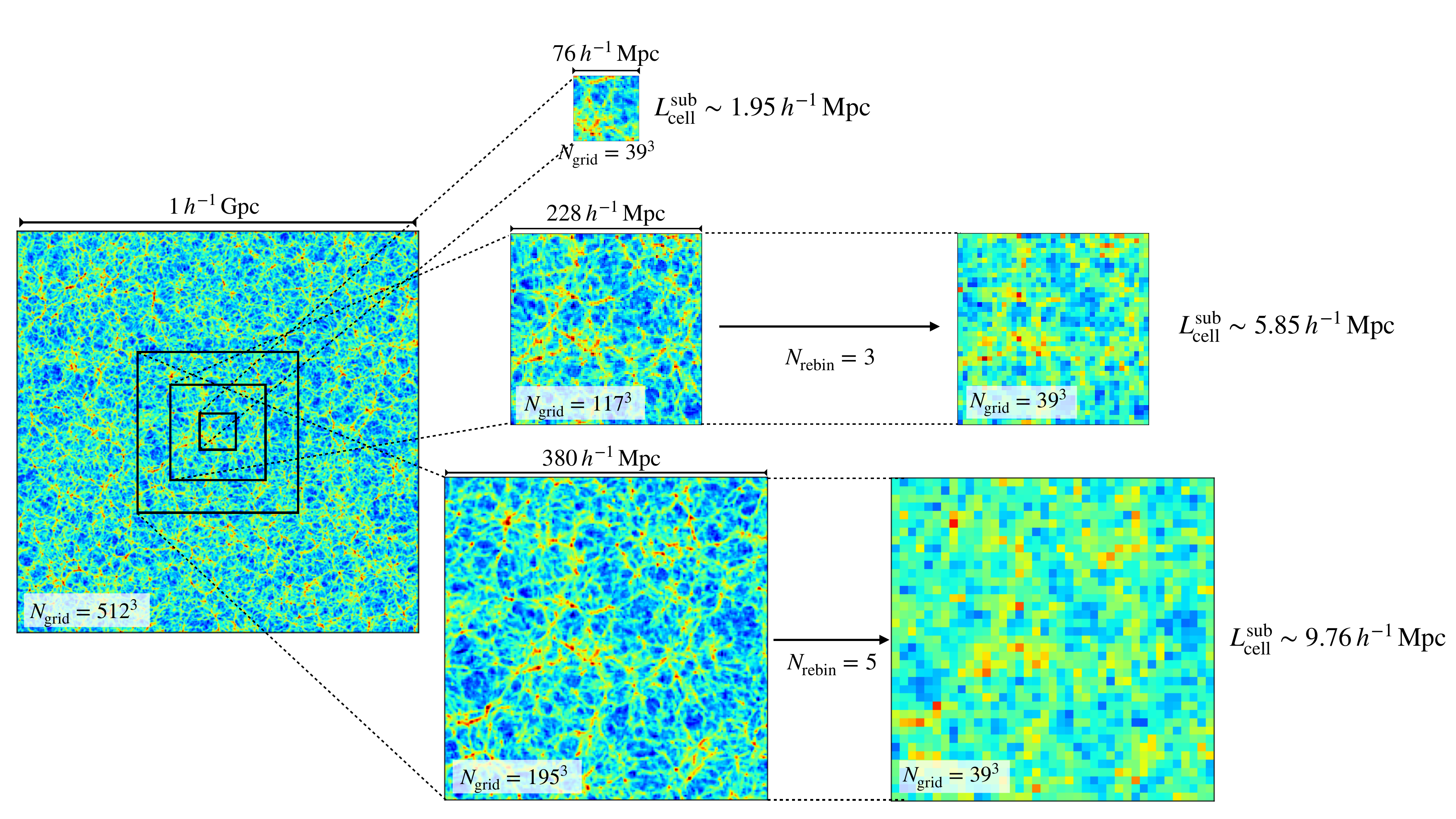}
    \caption{
        \revise{Generation of datasets for different sub-box configurations used as CNN inputs, as described in Section~\ref{subsec:dataset}. Shown here are three example sub-boxes with side lengths $\lsub \sim 76,228$ and $380\,\hmpc$, extracted from a parent box with $(L,\,\ngrid)=(1\,\hgpc,\,512^3)$, to estimate the target density $\delta_\mathrm{ini}$ at $(x,y,z)=(256,256,256)$. For clarity, we show 2D slices instead of full 3D boxes. First, a sub-box centred on the target position is extracted, retaining the same resolution as the parent box. Second, the grid of each sub-box is downsampled to $39^3$ cells by averaging over blocks of $(\nrebin)^3$. Thus, sub-boxes of different physical resolutions but identical grid sizes are constructed as CNN inputs. Alternative configurations employing different target resolutions are summarized in Table~\ref{tab:cnn_scales}.}
    }
    \label{fig:rebin}
\end{figure*}

\begin{table*}
 \caption{Network architecture of our new model; the dual-input 3D CNN. Each of the two input branches processes a $39^3$ voxel grid independently through seven convolutional layers with ReLU activations. To preserve spatial resolution, we apply padding in all convolutional layers except the final $1\times1\times1$ convolution. After feature extraction, the outputs from both branches are concatenated and passed through fully connected layers to produce a scalar prediction. Again, for the convolutional layers, the input and output tensors follow the format \texttt{[batch size, depth, height, width, channels]}. For the fully connected layers, the tensor shape is \texttt{[batch size, channels]}. We vary the batch size $\nbatch$ using values of 32 (iteration=[0,1]$\times10^5$), 128 (iteration=[1,3]$\times10^5$) and 512 (iteration=[3,5]$\times10^5$) during training.
 }
 \label{tab:cnn_dual}
 \centering
\begin{tabular}{lcccc}
\hline 
Layer & Kernel size & Output shape & Stride \\
\hline \hline
input$_1$, input$_2$ & None & ($\nbatch$, 39, 39, 39, 1), ($\nbatch$, 39, 39, 39, 1) & None  \\\hline
conv1, conv1' & (3, 3, 3) & ($\nbatch$, 20, 20, 20, 32), ($\nbatch$, 20, 20, 20, 32)  & (2, 2, 2)  \\\hline
conv2, conv2' & (3, 3, 3) & ($\nbatch$, 20, 20, 20, 32), ($\nbatch$, 20, 20, 20, 32) & (1, 1, 1)  \\\hline
conv3, conv3' & (3, 3, 3) & ($\nbatch$, 10, 10, 10, 64), ($\nbatch$, 10, 10, 10, 64) & (2, 2, 2)  \\\hline
conv4, conv4' & (3, 3, 3) & ($\nbatch$, 10, 10, 10, 64), ($\nbatch$, 10, 10, 10, 64) & (1, 1, 1)  \\\hline
conv5, conv5' & (3, 3, 3) & ($\nbatch$, 5, 5, 5, 128), ($\nbatch$, 5, 5, 5, 128) & (2, 2, 2)  \\\hline
conv6, conv6' & (3, 3, 3) & ($\nbatch$, 5, 5, 5, 128), ($\nbatch$, 5, 5, 5, 128) & (1, 1, 1)  \\\hline
conv7, conv7' & (1, 1, 1) & ($\nbatch$, 5, 5, 5, 128), ($\nbatch$, 5, 5, 5, 128) & (1, 1, 1)  \\\hline
flatten + concat & None & ($\nbatch$, 128000) & None  \\\hline
fully connected 1 & None & ($\nbatch$, 512) & None  \\\hline
fully connected 2 & None & ($\nbatch$, 128) & None \\\hline
fully connected 3 & None & ($\nbatch$, 1) & None  \\\hline\hline
\end{tabular}
\end{table*}

\subsection{Data set}
\label{subsec:dataset} 

We use the Indra simulations~\citep{Falck2021}, a suite of large-volume cosmological N-body simulations. Each of the simulations is computed with the WMAP7 cosmological parameters ($\Omega_m = 0.272$, $\Omega_\Lambda = 0.728$, $\Omega_b = 0.045$, $h = 0.704$, $\sigma_8 = 0.81$, $n_s = 0.967$;~\citet{Komatsu2011}) and different initial conditions using L-Gadget2~\citep{Springel2005}, each with $1024^3$ dark matter particles in a periodic cube of $1\hgpc$ on a side. For each simulation, the dark matter particles are projected onto a regular grid using the Cloud-In-Cell (CIC) scheme to construct the density field, which is then normalized by the mean density. As a result, all datasets used in this study are expressed in terms of the density contrast, defined as $\delta \equiv n_\mathrm{DM}/\bar{n}_\mathrm{DM} - 1$, where $n_\mathrm{DM}$ and $\bar{n}_\mathrm{DM}$ are the number density of dark matter and its mean value, respectively.
We define the snapshot at $z = 10$ as the initial condition and that at $z = 0$ as the final condition. 
The input of CNN is a cubic region of the final density field which grid size is $39^3$. The target value for training is the initial density of the central cell within the $39^3$ grids of the input extracted from the final condition.

As described in Section~\ref{subsec:cnn}, we vary the input sub-box sizes over a range of scales, 
from 38 to 380 $\hmpc$
as summarized in Table~\ref{tab:cnn_scales}. 
We begin by generating both input and target density fields within a $1~\hgpc$ simulation box using different grid resolutions ranging from $\ngrid=256^3$ to $1024^3$, resulting in varying cell sizes $\lcell^\mathrm{box}$. Following \citet{Mao2021}, the target field is smoothed using a Gaussian filter with a smoothing length of $3~\hmpc$. 
For the input field, we fix the number of grid points within each sub-box to $39^3$, and vary the the sub-box sizes by adjusting the cell size\revise{, as illustrated in Fig.~\ref{fig:rebin}.}
Specifically, we perform a "rebinning" operation in which each sub-box cell of size $\lcell^\mathrm{sub}$ is obtained by averaging over a block of $(\nrebin)^3$ cells from the full-resolution field, where $\lcell^\mathrm{sub} = \nrebin \times \lcell^\mathrm{box}$. This results in sub-boxes of fixed grid resolution $39^3$, but with ten different sub-box sizes $\lsub$, corresponding to ten different values of $\lcell^\mathrm{sub}$. This rebinning procedure reduces the resolution while preserving the overall large-scale structure of the density field, allowing us to probe the impact of input scale on reconstruction performance.
Note that the setup with $\lsub \sim 76~\hmpc$ 
is exactly same configurations used in \citet{Mao2021}.

For the dual-input network, we consider combinations of sub-boxes from $\lsub~[\hmpc] \sim $ \{76, 228, 380\}, resulting in three scale pairs: \{76, 228\}, \{228, 380\}, \{76, 380\}.
These configurations were chosen because they share the same grid resolution, $\ngrid = 512^3$, allowing sub-boxes at different scales to be aligned at a common central position. Although higher-resolution cases with $\ngrid = 1024^3$ would also permit such alignment, they incur computational costs beyond our available resources. 
as seen in Table~\ref{tab:cnn_scales}
Each sub-box in a pair, sharing the same central position, is fed into a separate CNN branch to extract scale-specific features, which are then concatenated and passed to the fully connected layers.


The training set is sampled such that sub-boxes are extracted at intervals of $\ngrid/32$ grid cells along each spatial dimension, leading to $32^3=32,768$ sub-boxes in a single simulation. We use eight independent realisations for constructing the entire training sample, which totals 262,144 sub-boxes.
To augment the training data, we apply six unique rotations and eight axis reflections to each sub-box, increasing the dataset size by a factor of 48. For validation, 4,096 sub-boxes are randomly selected from an independent simulation without augmentation to monitor potential overfitting during training. We note that we never use the validation set for the optimization.

In addition to the training and validation datasets, an independent test set is employed to evaluate the final performance of the trained model. Unlike in training and validation, the test data are not used in the loss computation during optimization.
All results shown in Section~\ref{sec:results} are based on a single test simulation, which shares the same cosmological parameters as the training and validation sets but differs in its initial conditions.
We generate the test set in the same way as the training set, but their sub-boxes are extracted at no intervals, leading to $\ngrid^3$ sub-boxes in a single simulation without any data augmentation.

\subsection{Training and Validation}
\label{subsec:train}
In machine learning models, hyper-parameters are external configuration variables -- such as learning rate, batch size, and network depth -- that are not learned from the data but significantly influence training performance and outcomes. In this study, we 
follow
the set of hyper-parameters fine-tuned by \citet{Mao2021}, and therefore do not perform additional hyper-parameter optimization.

Our implementation is based on the PyTorch~\citep{Paszke2019} deep learning framework. Network weights are initialized using the Xavier initialization scheme~\citep{Glorot2010}. We employ the Adam optimizer \citep{Kingma2014}, 
which combines the advantages of adaptive learning rates and momentum to achieve efficient and robust convergence. 

In our training procedure, the dataset is divided into \textit{batches}, where each batch consists of a subset of the full training set. The number of samples in each batch is referred to as the batch size, denoted as $\nbatch$, leading to the total number of batches equal to 262,144$\,\times\,48/\nbatch$.
The loss function 
is the mean squared error (MSE), defined as
\begin{align}
\label{eq:mse}
\mathrm{MSE} = \frac{1}{\nbatch}\sum_{k=1}^{\nbatch}\left(f(\delta_\mathrm{fin}^k;\bm{\theta})-\delta_\mathrm{ini}^k\right)^2,
\end{align}
where $\delta_\mathrm{fin}^k$ represents the input density field of size $39^3$ for batch element $k \in \nbatch$, and $\delta_\mathrm{ini}^k$ denotes the corresponding scalar value of the target initial density.
The MSE quantifies the average squared difference between the predicted values $f(\delta_\mathrm{fin}^k;\bm{\theta})$ with a parameter set $\bm{\theta}$ and the target values $\delta_\mathrm{ini}^k$ for each batch. In supervised learning, the loss function plays a central role in training by quantifying the discrepancy between the network's predictions and the ground truth. It serves as the primary indicator of model performance and is used to guide the optimization of trainable parameters through gradient-based updates.

We define the \textit{iteration} as a single update step of the model parameters using a batch of training data. During each iteration, the CNN processes a batch of input and target density fields, computes the loss function, and updates the model weights via backpropagation using the optimizer. Following the strategy of \citet{Mao2021}, we fix the learning rate to 
$10^{-4}$
and vary the batch size $\nbatch$ 
as 32 during the number of iterations $[0,1]\times10^5$, 128 ($[1, 3]\times10^5$) and 512 ($[3,5]\times10^5$).
To monitor generalization and detect overfitting, we track the loss on both the training and validation sets. The validation set, which is not used for model updates, provides an unbiased estimate of performance. Overfitting is indicated when training loss decreases while validation loss increases.

\section{Results}
\label{sec:results}
First, in Section~\ref{subsec:result_optimal_scale}, we present the results from the variety of input sub-box sizes in a range $\lsub\sim38-380\hmpc$, as shown in Table~\ref{tab:cnn_scales}. Second, in Section~\ref{subsec:result_multi}, we display the performance of our new network model, dual-input 3D CNN, using the input sub-boxes of $\lsub\sim76\hmpc$ and $\lsub\sim228\hmpc$.

To evaluate the precision of our reconstruction, we 
show the loss function curve and, 
the probability distribution function (PDF) of the reconstructed fields. In addition, we assess the accuracy of the reconstruction by examining the transfer function and the correlation coefficient between the true initial conditions and the network outputs.

\begin{figure*}
 \begin{tabular}{c|c}
\includegraphics[width=0.45\linewidth]{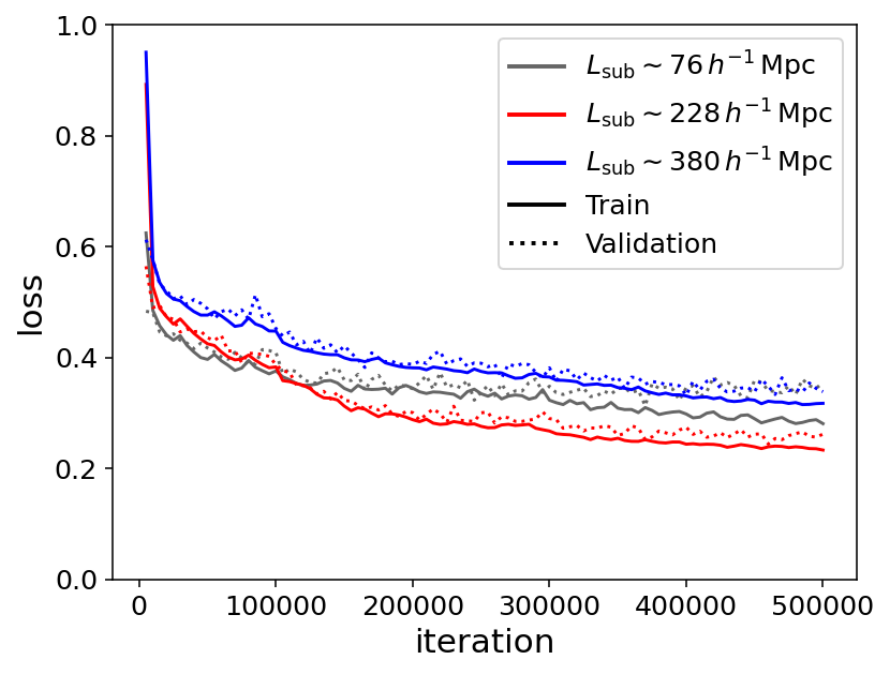} &
\includegraphics[width=0.45\linewidth]{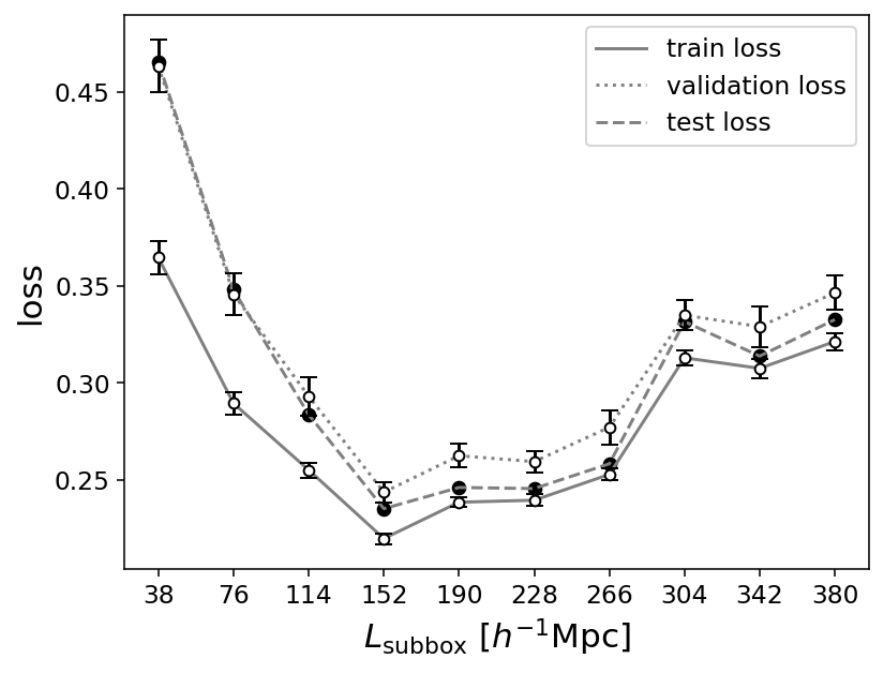}\\
 \end{tabular}
 \caption{\textit{Left} : Loss functions defined in equation~(\ref{eq:loss_function}). Solid and dotted lines represent the train and validation loss, respectively. Results are shown for three representative sub-box sizes: $\lsub\sim76\hmpc$ (matching the setup of \citet{Mao2021}; gray), \revise{$\lsub\sim228\hmpc$} (intermediate size; red) and $\lsub\sim380\hmpc$ (large size; blue). \textit{Right} : Train and validation loss as a function of sub-box size $\lsub$. White circles and error-bars denote the mean and standard deviation of the loss over $[4,5]\times10^5$ iterations. Black points indicate loss values computed from all grid points in a test simulation.
}
 \label{fig:loss}
\end{figure*}

\begin{figure*}
 \begin{tabular}{c|c}
\includegraphics[width=0.45\linewidth]{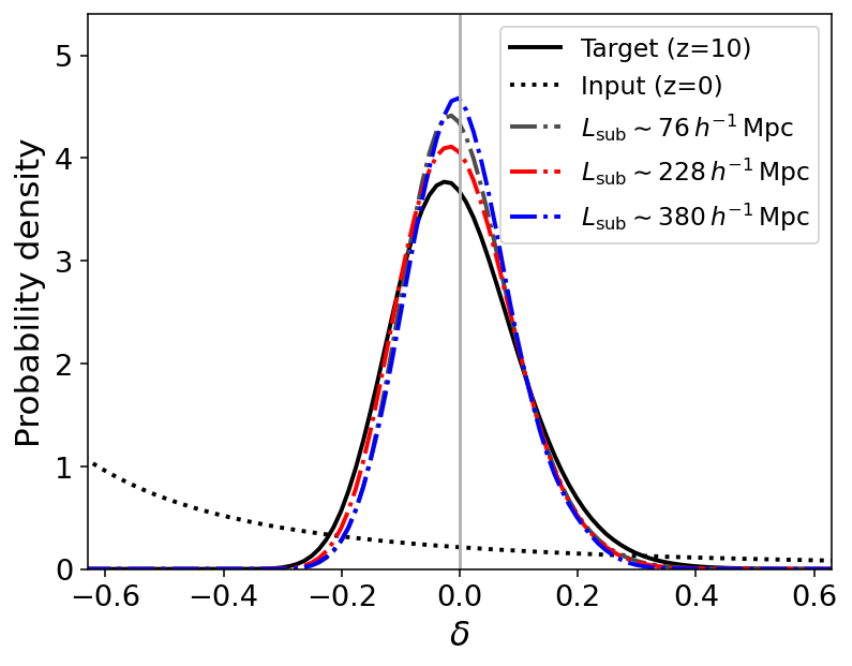} &
\includegraphics[width=0.45\linewidth]{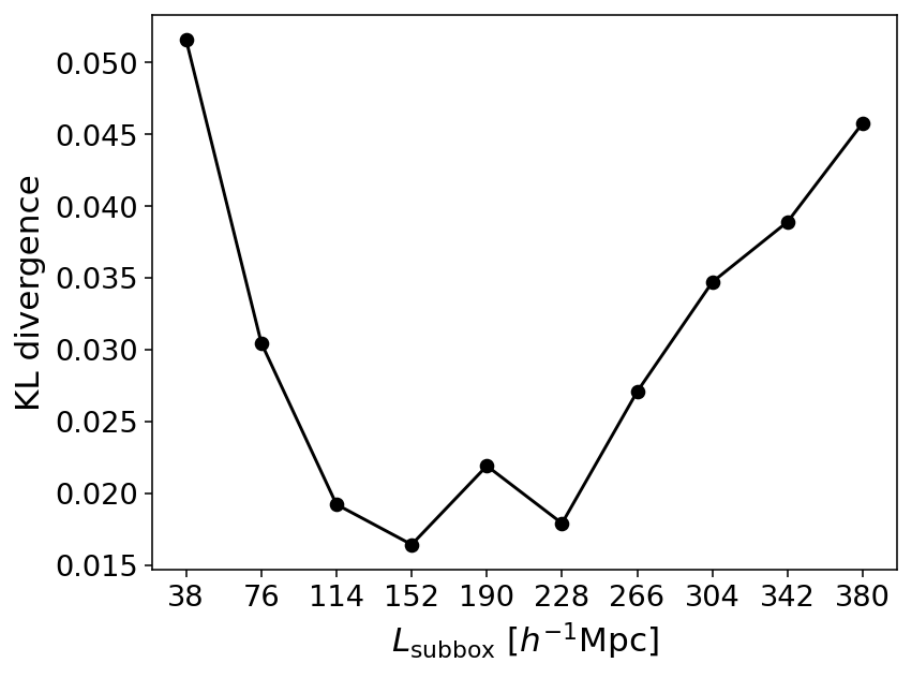}\\
 \end{tabular}
 \caption{\textit{Left} : PDFs of the density fluctuation $\delta$. Solid, dotted and dash-dotted lines denote PDFs of the target $\dini$, input $\dfin$ and output $\drec$ density distributions, respectively. Results are shown for three representative sub-box sizes: $\lsub\sim76,\,\revise{228}\,\mathrm{and}\,380\hmpc$. \textit{Right} : KL divergence as a function of sub-box size $\lsub$. KL divergence between the predicted and target initial density distributions, defined in equation~(\ref{eq:kl_divergence}). 
}
 \label{fig:pdfs}
\end{figure*}

\subsection{Varying scales of input grids}
\label{subsec:result_optimal_scale}

\begin{figure*}
 \begin{tabular}{c|c}
\includegraphics[width=0.45\linewidth]{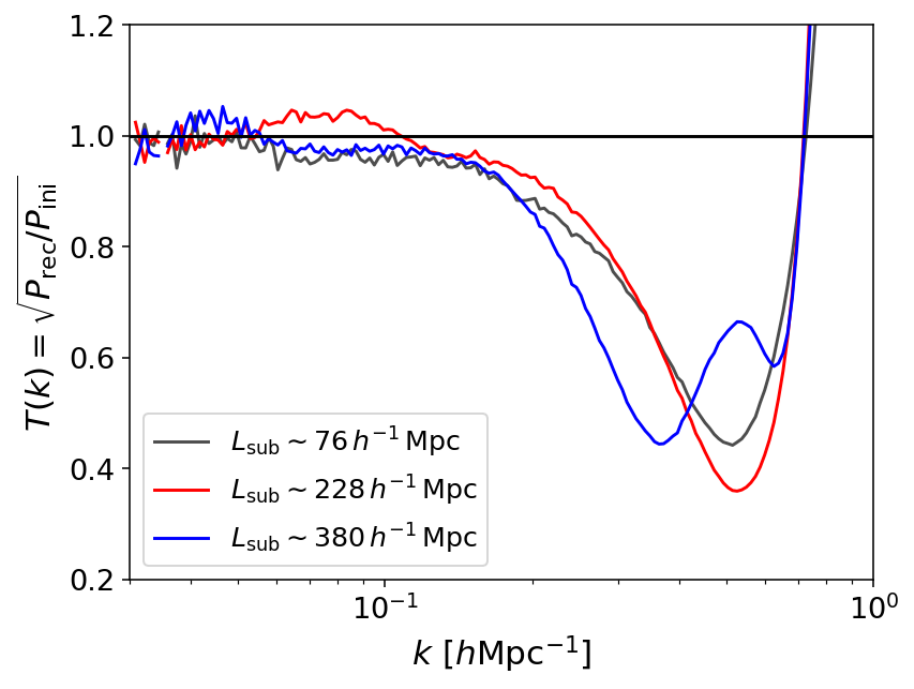} &
\includegraphics[width=0.45\linewidth]{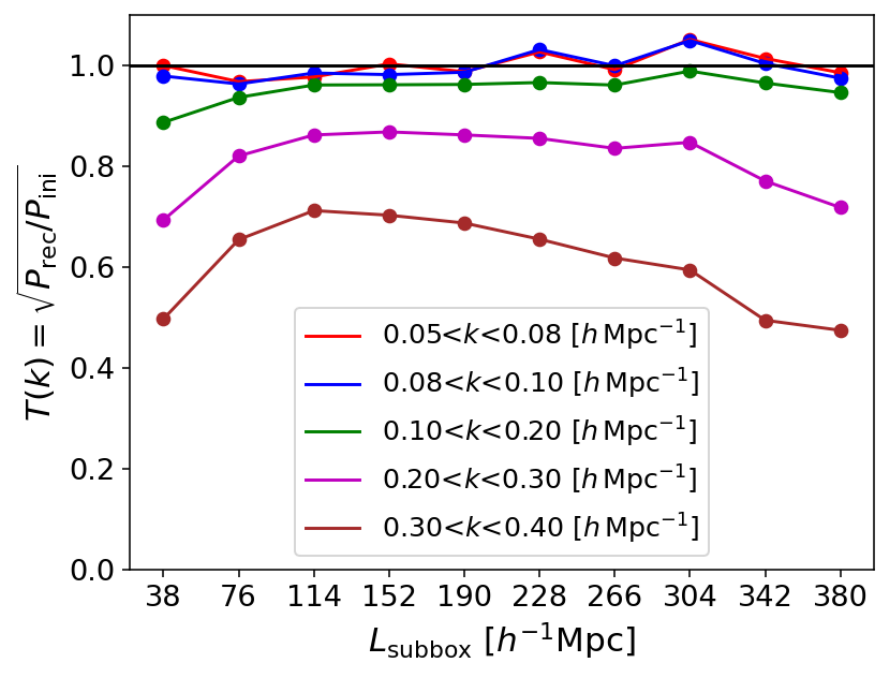}\\
 \end{tabular}
 \caption{\textit{Left} : Transfer function between the power spectrum of the reconstructed and target initial density fields, defined in equation~\ref{eq:transfer}. Results are shown for three representative sub-box sizes: $\lsub\sim76,\,\revise{228}\,\mathrm{and}\,380\hmpc$ from a single simulation. \textit{Right} : Transfer function averaged over five wavenumber bins, as a function of sub-box size $\lsub$. 
}
 \label{fig:transfer}
\end{figure*}

\begin{figure*}
 \begin{tabular}{c|c}
\includegraphics[width=0.45\linewidth]{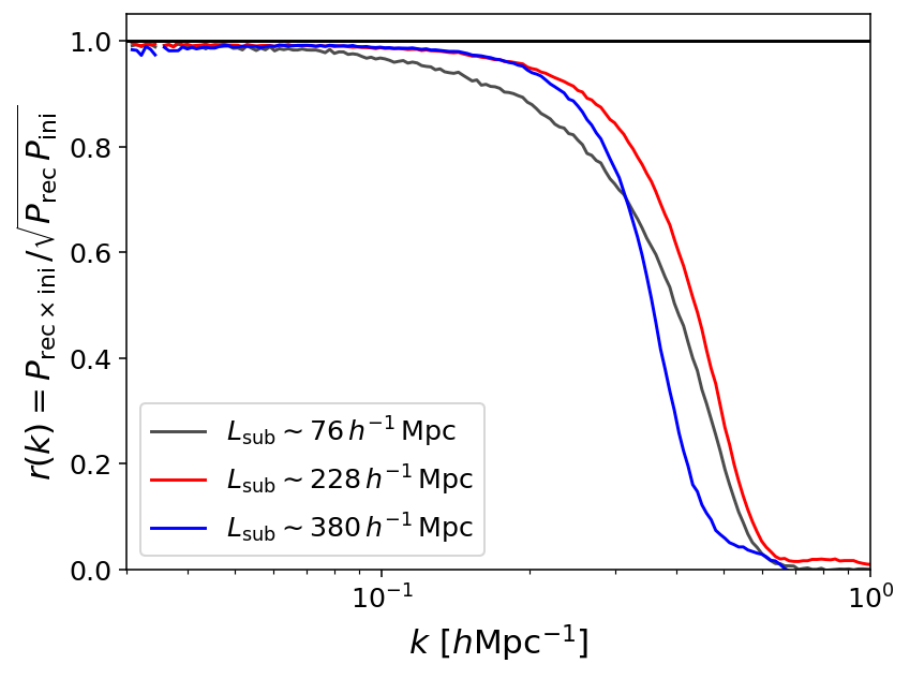} &
\includegraphics[width=0.45\linewidth]{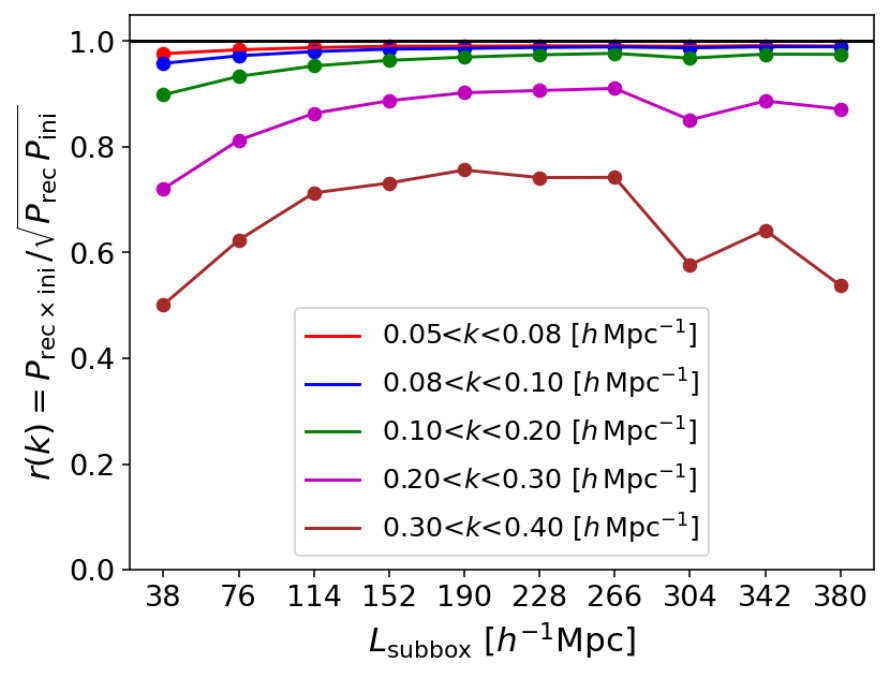}\\
 \end{tabular}
 \caption{\textit{Left} : Correlation coefficient between the power spectrum of the reconstructed and target initial density fields, defined in equation~\ref{eq:correlation}. Results are shown for three representative sub-box sizes: $\lsub\sim76,\,\revise{228}\,\mathrm{and}\,380\hmpc$ from a single simulation. \textit{Right} : Correlation coefficient averaged over five wavenumber bins, as a function of sub-box size $\lsub$. 
}
 \label{fig:corref}
\end{figure*}

In Fig.~\ref{fig:loss}, we present the 
loss function normalized by the variance of the initial density field, and defined as:
\begin{align}
\label{eq:loss_function}
\frac{\mathrm{MSE}}{\sigma^2} = \frac{\sum_k\left(f(\delta_\mathrm{fin}^k;\bm{\theta})-\delta_\mathrm{ini}^k\right)^2}{\sum_k\left(\bar{\delta}_\mathrm{ini}-\delta_\mathrm{ini}^k\right)^2},
\end{align}
where $f(\delta_\mathrm{fin}^k;\bm{\theta})$ denotes the network output, $\delta_\mathrm{ini}^k$ the true initial density, and $\bar{\delta}_\mathrm{ini}$ the mean of the initial density field. This normalization ensures that the loss equals unity when the network output is simply the mean value, regardless of the input.
The left panel of Fig.\ref{fig:loss} shows normalized losses of the training and validation for three 
different sub-box sizes. 
\revise{For a fair comparison, we present three cases with identical target resolutions ($\lsub$ = 76, 228 and 380$\hmpc$).}
We observe that although the training loss is still decreasing, the validation loss is almost converged.
Therefore, we stop learning and use 
models at iteration\,=\,500{,}000 for reconstruction of the test set. In the following, we refer to the output from trained model as the reconstructed density $\drec$.
\revise{It is also found that differences in the size of $\lsub$ lead to variations in the learning efficiency of the CNN.}
Among the three, the $\lsub \sim \revise{228}~\hmpc$ case exhibits the fastest convergence and reaches the lowest final loss, suggesting better reconstruction performance at this scale.

The right panel of Fig.~\ref{fig:loss} shows the mean and standard deviation of the normalized training and validation losses, averaged over iterations from $[4,5]\times10^5$, for each output corresponding to the ten different sub-box sizes $\lsub$.
The test loss is calculated by the trained model with a simulation run independent of the train and validation. 
We see that the training loss decreases monotonically as the input scale increases up to $\lsub\sim152\hmpc$, reaching a minimum at $\lsub\sim152\hmpc$.
A similar trend is observed in both the validation and test losses. 

In addition to the MSE/$\sigma^2$, we introduce another metric to quantify the goodness of the reconstruction. Figure~\ref{fig:pdfs} presents the probability density functions (PDFs; left panel) and the corresponding Kullback–Leibler (KL) divergences (right panel) between the predicted and true initial density fields. The KL divergence is defined as
\begin{align}
\label{eq:kl_divergence}
{D_{\rm KL}(p \| q)=\sum_{i=1}^{N} p\left(x_{i}\right) \cdot\left[\log p\left(x_{i}\right)-\log q\left(x_{i}\right)\right],
}
\end{align}
where $p(x_i)$ and $q(x_i)$ denote the discrete PDFs of the target (true initial density, $\dini$) and the network output (reconstructed density, $\drec$), respectively. We use $D_{\rm KL}(p \| q)$ to quantify the similarity between the two distributions: a smaller value indicates greater similarity. 

The left panel of Fig.~\ref{fig:pdfs} shows the probability density functions (PDFs) of the target initial density field $\dini$, the input final density field  $\dfin$ and the predicted output $\drec$  for three representative cases. In all examples, the predicted PDFs are noticeably closer to the target than those of the input, indicating that the network effectively extracts information about the initial density field from the final condition. However, the predicted distributions exhibit a slightly reduced variance compared to the true initial density field, suggesting that the learned CNN model may "pull back" the density fluctuations too aggressively during the reconstruction process.

The right panel of Fig.~\ref{fig:pdfs} presents the 
KL divergence between the predicted and target initial density distributions as a function of the sub-box size $\lsub$. As shown, the KL divergence decreases significantly as $\lsub$ increases from $38\hmpc$ to $152\hmpc$, reaching a minimum at $\lsub\sim152\hmpc$. 
These results highlight the importance of 
the scales correlated with the
recovery of the initial density field.

Now we are looking into the reconstruction accuracy as a function of scales. To do so, we work on the Fourier space rather than map basis comparison.
The left panel of Fig.~\ref{fig:transfer} shows the transfer function which provides a measure of the agreement between the power spectra of the reconstructed and true initial density fields, 
\begin{align}
\label{eq:transfer}
    T(k)=\sqrt{\frac{P_\mathrm{rec}(k)}{P_\mathrm{ini}(k)}},
\end{align}
where $P_\mathrm{rec}(k)$ and $P_\mathrm{ini}(k)$ denote the three-dimensional power spectra of the reconstructed field $\drec$ and the true initial field $\dini$, respectively. On large scales ($k\lesssim0.2\ihmpc$), the transfer functions 
fluctuate within the range $0.9\lesssim T(k) \lesssim 1.1$. On smaller scales ($0.2\ihmpc\lesssim k \lesssim0.4\ihmpc$), the $\lsub\sim\revise{228}\hmpc$ 
reconstruction exhibits a slightly better performance over other two cases.

In the right panel of Fig.~\ref{fig:transfer}, we show the transfer functions averaged over five wavenumber bins: ($0.05<k<0.08$, $0.08<k<0.10$, $0.10<k<0.20$, $0.20<k<0.30$ and $0.30<k<0.40 \ihmpc$), as a function of the sub-box size $\lsub$. On the largest scales ($0.05<k<0.08\ihmpc$), the transfer function is close to unity across all $\lsub$, indicating that the reconstruction well reproduce the large-scale power. This 
is also observed
in the second bin ($0.08<k<0.10\ihmpc$) where $T(k)$ remains within $\sim2-5\%$ around unity and shows little variation with sub-box size. In the intermediate bin ($0.10<k<0.20\ihmpc$), the transfer function values lie 
around $T(k)\sim0.96$, indicating mild suppression of power on these scales. The dependence on $\lsub$ is weak, though slightly higher values are seen for larger sub-boxes.
More prominent differences emerge on smaller scales. In the bin $0.20<k<0.30\ihmpc$, the transfer function improves significantly with increasing $\lsub$ up to $152\hmpc$, peaking near $T(k)\sim0.87$, and then declines toward both smaller and larger sub-box sizes. 
This trend suggests that $\lsub\sim152\hmpc$ represents an optimal trade-off between preserving local resolution and incorporating sufficient large-scale information within this wavenumber range.
In the smallest-scale bin ($0.30<k<0.40 \ihmpc$), the transfer function decreases more rapidly with $\lsub$, reflecting the loss of small-scale information due to limited effective resolution. The optimal performance is again seen near with $\lsub\sim114-190\hmpc$ consistently provides the best performance across all $k$-bins.

While the transfer function quantifies the reconstruction accuracy in terms of the amplitude of fluctuations at each scale, it does not capture information about the phase. Therefore, it is important to further investigate how well the phase information of the fluctuations is reproduced across different scales.
We quantify the cross-correlation between the reconstructed and initial density fields in Fourier space using the correlation coefficient, 
\begin{align}
\label{eq:correlation}
    r(k)=\frac{P_\mathrm{rec\times ini}(k)}{\sqrt{P_\mathrm{rec}(k)P_\mathrm{ini}(k)}}~,
\end{align}
where $P_\mathrm{rec\times ini}(k)$ is the cross-power spectrum between the reconstructed and initial fields, and $P_\mathrm{rec}(k)$ and $P_\mathrm{ini}(k)$ are their respective auto-power spectra. As shown in the left panel of Fig.~\ref{fig:corref}, the correlation coefficient remains close to unity ($r(k)\gtrsim0.99$) on large scales ($k\lesssim0.05\ihmpc$) and gradually declines toward smaller scales.
The transition scale where drops below 0.9 is around $k\sim0.2\ihmpc$, beyond which the correlation rapidly decreases. This behavior confirms that the reconstruction is highly accurate on large scales but progressively less reliable on small scales.

In the right panel of Fig.~\ref{fig:corref}, we evaluated the correlation coefficient $r(k)$ as a function of sub-box size $\lsub$ averaged over the five wavenumber bins. On the large scales ($k<0.1\ihmpc$), the correlation coefficients are high across all sub-box sizes, with $r(k)>0.95$ even for the smallest sub-boxes. In the intermediate bin ($0.10<k<0.20\ihmpc$), the correlation coefficients are slightly below unity, around $r(k)\sim0.95$. On 
smaller scales ($0.20<k<0.40\ihmpc$), the correlation coefficients are smaller than 0.9 but they 
have peak around $\lsub \sim 190-226\hmpc$.
Again, the optimal $\lsub$ can be found in 190-226$\hmpc$, which is consistent with other statistics.

\begin{figure}
\centering
    \centering
    \includegraphics[width=0.9\columnwidth]{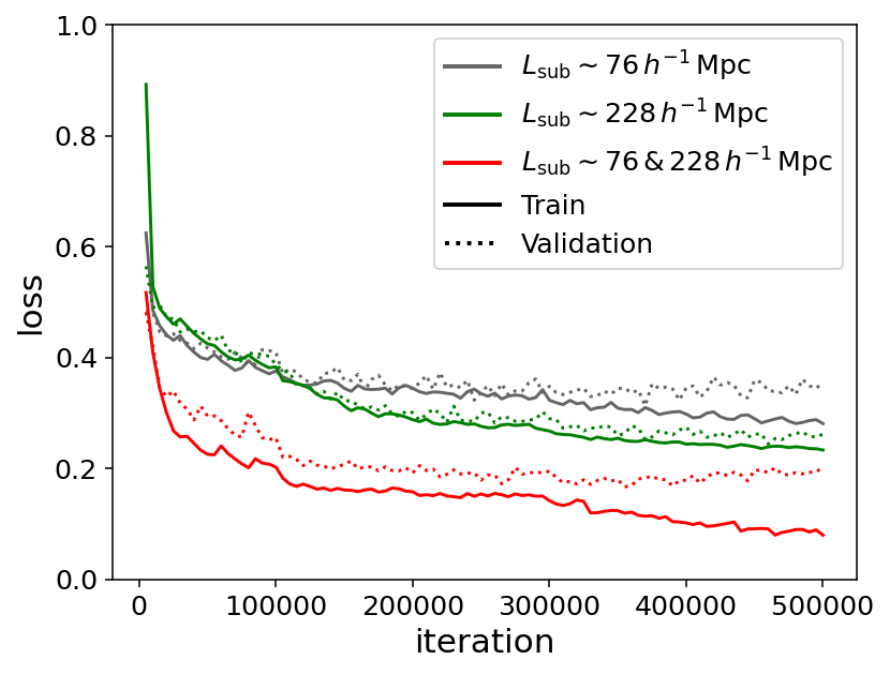}
    \caption{Loss functions as defined in equation~(\ref{eq:loss_function}). Results are shown for two representative sub-box sizes: $\lsub\sim76\hmpc$ (gray) and $\lsub\sim228\hmpc$ (green), as well as for dual-input model using $\lsub\sim\{76,228\}\hmpc$ as inputs (red).}
    \label{fig:loss_multi}
\end{figure}
\begin{figure}
    \centering
    \includegraphics[width=0.9\columnwidth]{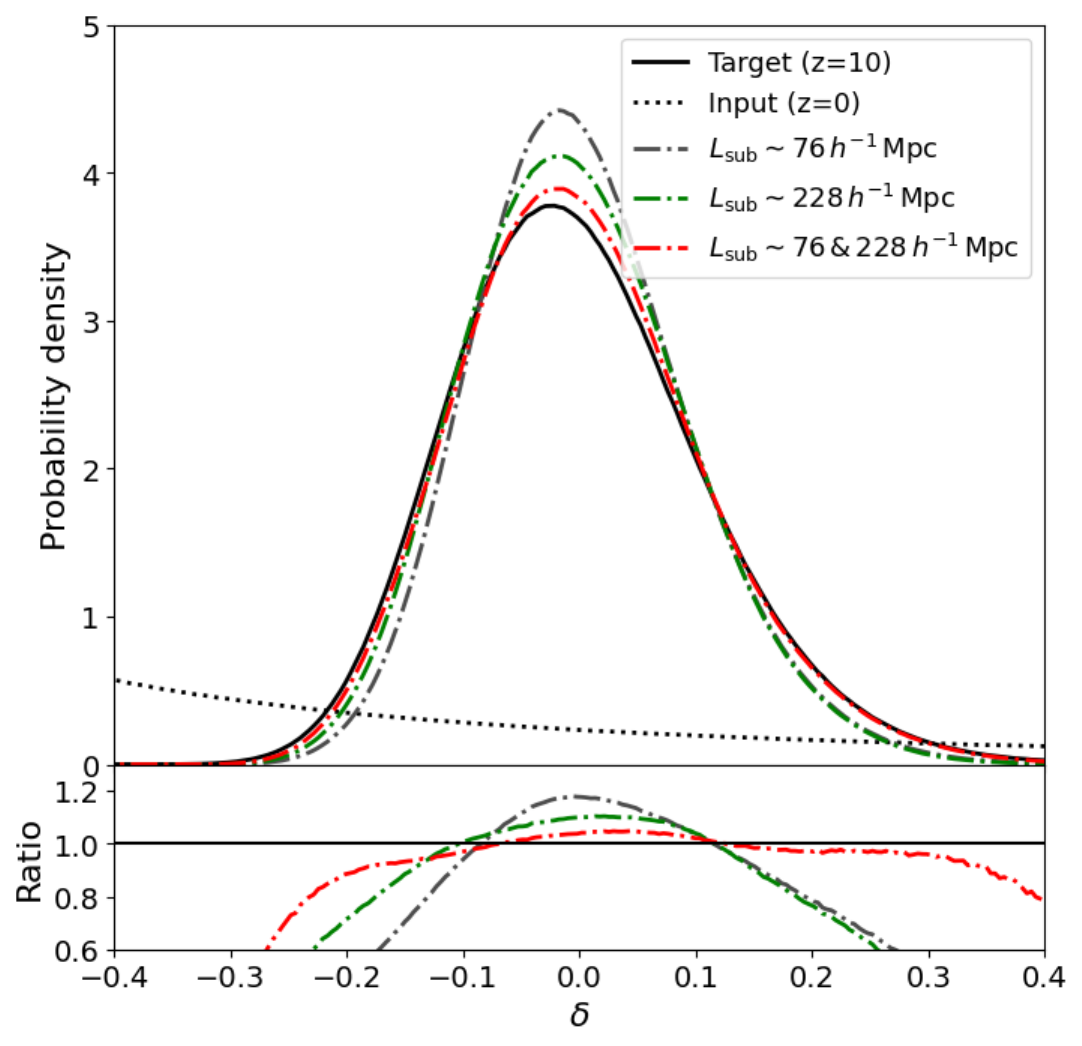}
    \caption{PDFs of the density fluctuation $\delta$ and ratio between a target and three outputs. Results are shown for two representative sub-box sizes: $\lsub\sim76\,\mathrm{and}\,228\hmpc$, as well as for dual-input model ($\lsub\sim\{76,228\}\hmpc$).}
    \label{fig:pdf_multi}
\end{figure}

\begin{figure}
\centering
    \centering
    \includegraphics[width=0.9\columnwidth]{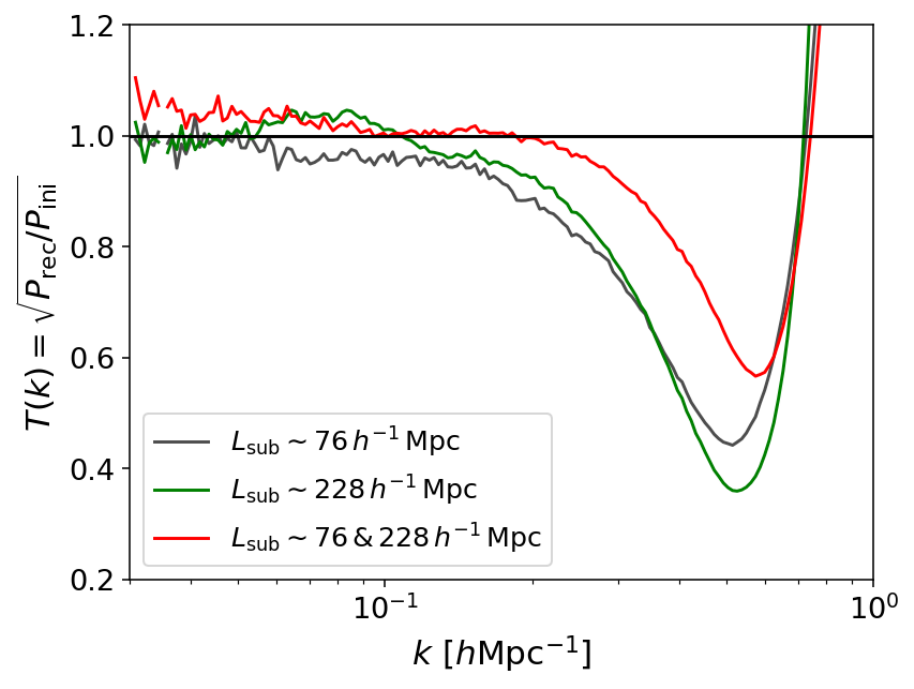}
    \caption{Transfer function between the power spectrum of the reconstructed and target initial density fields, defined in equation~\ref{eq:transfer}. Results are shown for two representative sub-box sizes: $\lsub\sim76\,\mathrm{and}\,228\hmpc$, as well as for dual-input model ($\lsub\sim\{76,228\}\hmpc$).}
    \label{fig:tk_multi}
\end{figure}
\begin{figure}
    \centering
    \includegraphics[width=0.9\columnwidth]{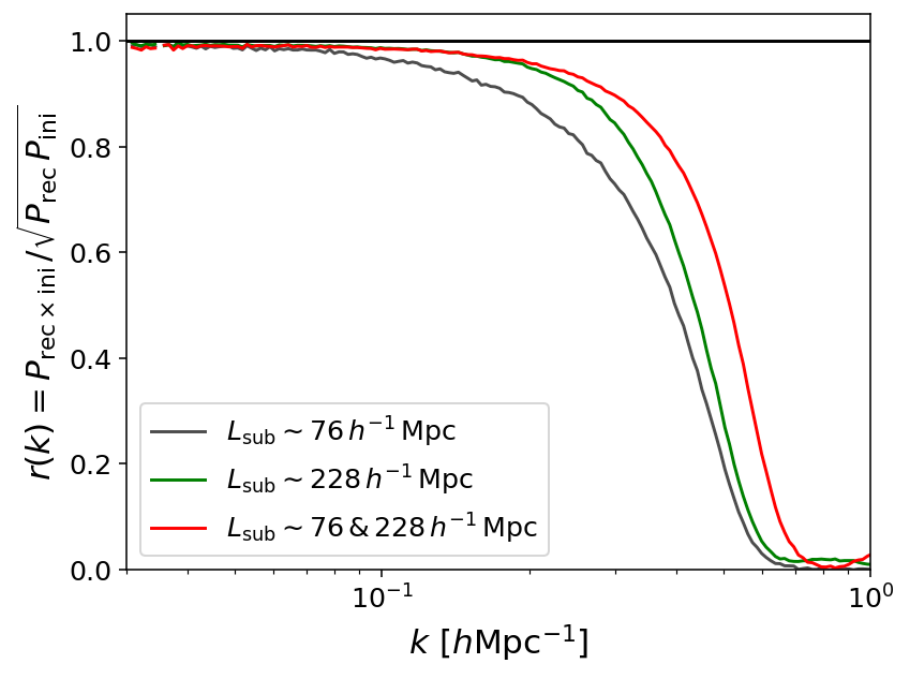}
    \caption{Correlation coefficient between the power spectrum of the reconstructed and target initial density fields, defined in equation~\ref{eq:correlation}. Results are shown for two representative sub-box sizes: $\lsub\sim76\,\mathrm{and}\,228\hmpc$, as well as for dual-input model ($\lsub\sim\{76,228\}\hmpc$).}
    \label{fig:rk_multi}
\end{figure}

\begin{table*}
 \caption{
Comparison of reconstruction accuracy between the best-performing single-input model and the dual-input model. The dual-input model uses two sub-box scales ($\lsub\sim76$ and $228~\hmpc$), while the single-input results shown correspond to the optimal sub-box sizes for each metric. Reported values include the mean and standard deviation of the validation loss over iterations $[4,5]\times10^5$, the KL divergence of the predicted and true initial density PDFs, and the averaged transfer function $T(k)$ and correlation coefficient $r(k)$ over three $k$-bins. Bold values indicate the best performance for each metric.}

 \label{tab:accuracy}
 \centering
\begin{tabular}{lcc}
\hline 
{} & \begin{tabular}{c} single-input (best $\lsub$) \end{tabular} & \begin{tabular}{c} dual-input ($\lsub\sim\{76,228\}\hmpc$) \end{tabular} \\\hline
\begin{tabular}{l} validation loss \end{tabular} & \begin{tabular}{c} $0.244\pm0.005$  ($\lsub\sim 152\hmpc$) \end{tabular} & \begin{tabular}{c}$\bm{0.190}\pm0.007$ \end{tabular}\\\hline
\begin{tabular}{l}KL divergence \end{tabular} & \begin{tabular}{c} 0.0164  ($\lsub\sim 152\hmpc$) \end{tabular} & \begin{tabular}{c}$\bm{0.0027}$\end{tabular}\\\hline


\begin{tabular}{l} $T(0.10<k<0.20\,[\ihmpc])$\\$T(0.20<k<0.30\,[\ihmpc])$\\$T(0.30<k<0.40\,[\ihmpc])$ \end{tabular} & \begin{tabular}{c} 0.988  ($\lsub\sim 304\hmpc$)  \\ 0.868  ($\lsub\sim 152\hmpc$) \\  0.712  ($\lsub\sim 114\hmpc$)  \end{tabular} & \begin{tabular}{c} $\bm{1.005}$\\ $\bm{0.963}$ \\ $\bm{0.861}$\end{tabular}\\\hline
\begin{tabular}{l} $r(0.10<k<0.20\,[\ihmpc])$\\$r(0.20<k<0.30\,[\ihmpc])$\\$r(0.30<k<0.40\,[\ihmpc])$ \end{tabular} & \begin{tabular}{c}  $\bm{0.977}$ ($\lsub\sim 266\hmpc$)  \\0.911  ($\lsub\sim 266\hmpc$) \\ 0.756  ($\lsub\sim 190\hmpc$) \end{tabular} & \begin{tabular}{c} 0.975\\$\bm{0.932}$ \\$\bm{0.842}$\end{tabular}\\\hline\hline

\end{tabular}
\end{table*}

\subsection{Dual input 3D CNN model}
\label{subsec:result_multi}
In this section, we present the loss function, PDFs, transfer function, and correlation coefficient obtained from the output of the dual-input 3D CNN model, which uses sub-boxes of size $\lsub\sim 76$ and $228 \hmpc$ as inputs. These results are compared with those from single-input models and summarised in Table~\ref{tab:accuracy}.
Results from the other pairs of dual-input model, $\lsub~[\hmpc]\sim$ \{228, 380\} and \{76, 380\}, are presented in Appendix~\ref{sec:app_multi}.

Figure~\ref{fig:loss_multi} shows the loss function. The red line represents  the results of our dual-input model. In the case of $\lsub\sim\{76,228\}\hmpc$, both the training and validation losses decrease smoothly as the number of iterations increases, up to approximately $300{,}000$. We see that our dual-input model learns the dataset more efficiently than the single-input models (shown in gray and green). 
After 300,000 iterations, we observe a slight increase in the validation loss, which suggests an onset of over-fitting. Therefore, we manually apply an early stopping at iteration 300,000.

In Fig.~\ref{fig:pdf_multi}, we present the PDFs of the reconstructed density perturbation for the dual-input model, alongside two representative examples from single-input models. The predicted PDF of the dual-input model shows significantly better agreement with the target than those of the single-input models, indicating that our new network architecture with two different input sub-box more effectively extracts information about the initial density field from the final condition. Furthermore, the dual-input model mitigates the tendency of single-input models to excessively "pull back" density fluctuations. Quantitatively, the KL divergence of the dual-input model is 0.0027, approximately $80\%$ smaller than the value of 0.0164 obtained from the single-input model with $\lsub\sim152\hmpc$, as summarised in Table~\ref{tab:accuracy}. 

Figure~\ref{fig:tk_multi} presents the transfer function between the power spectra of the reconstructed and true initial density fields. On large scales  ($k\lesssim0.2\ihmpc$), 
both single- and dual-input model
exhibit transfer functions fluctuating within the range $0.9 \lesssim T(k) \lesssim 1.1$. On smaller scales ($0.2\ihmpc\lesssim k \lesssim0.5\ihmpc$), the dual-input model maintains a transfer function closer to unity than the single-input models, indicating improved retention of small-scale power in the reconstruction. Specifically, the transfer function averaged over the ranges $0.2\ihmpc< k <0.3\ihmpc$ and $0.3\ihmpc< k <0.4\ihmpc$ reaches 0.963 and 0.861, which are approximately $10\%$ and $20\%$ higher, respectively, than the corresponding values of 0.868 and 0.712 obtained from the single-input model, as summarised in Table~\ref{tab:accuracy}.

Figure~\ref{fig:rk_multi} shows the correlation coefficient between the power spectra of the reconstructed and true initial density fields. 
On small scales ($0.1\ihmpc\lesssim k $), the dual-input model 
shows higher correlation coefficients, suggesting better preservation of small-scale cross-correlations. Quantitatively, the correlation coefficient averaged over the ranges $0.2\ihmpc< k <0.3\ihmpc$ and $0.3\ihmpc< k <0.4\ihmpc$  is 0.932 and 0.842, representing improvements of approximately $2\%$ and $11\%$ over the values of 0.911 and 0.756 from the single-input model, respectively, as listed in Table~\ref{tab:accuracy}.

\begin{figure*}
 \begin{tabular}{c|c}
\includegraphics[width=0.45\linewidth]{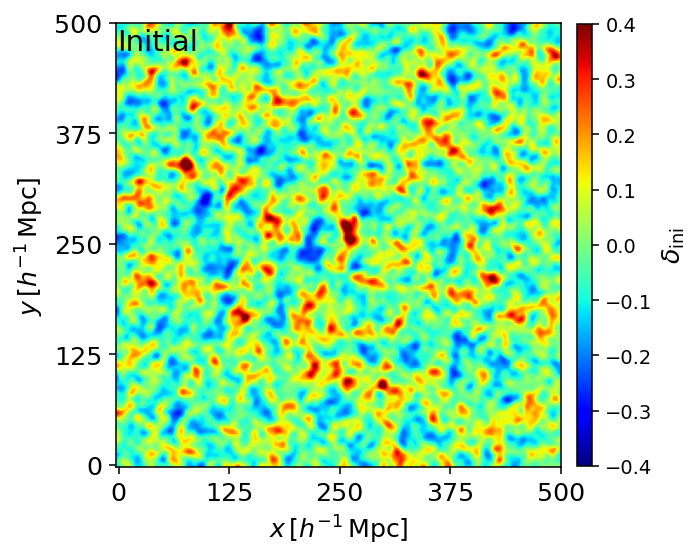} &
\includegraphics[width=0.45\linewidth]{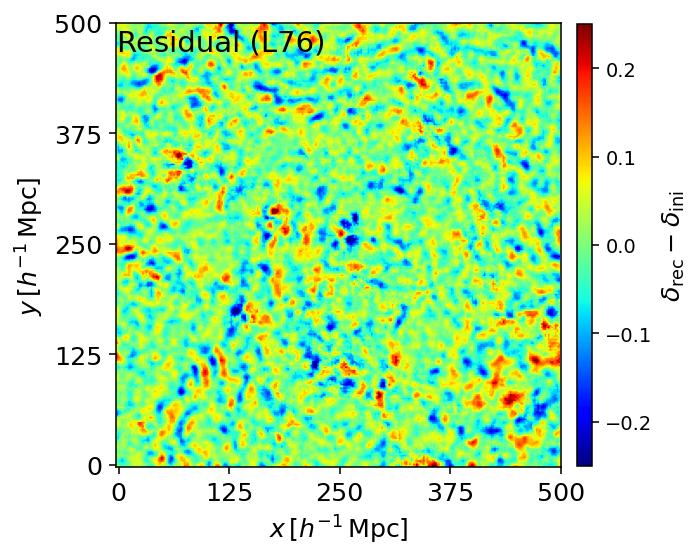}\\
\includegraphics[width=0.45\linewidth]{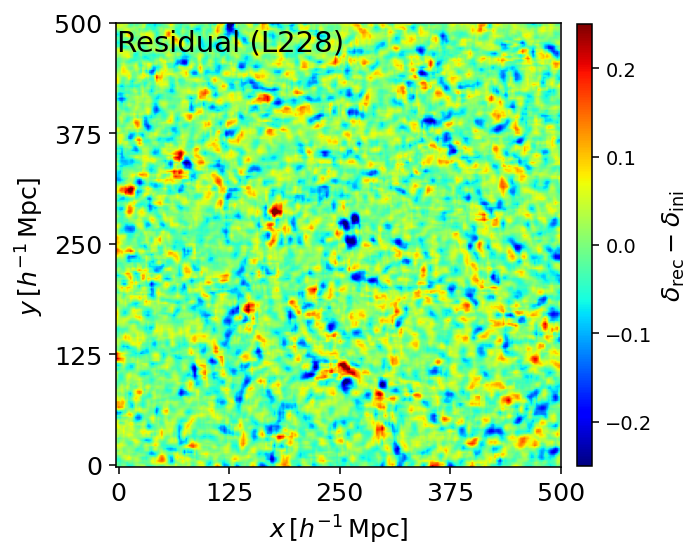}&
\includegraphics[width=0.45\linewidth]{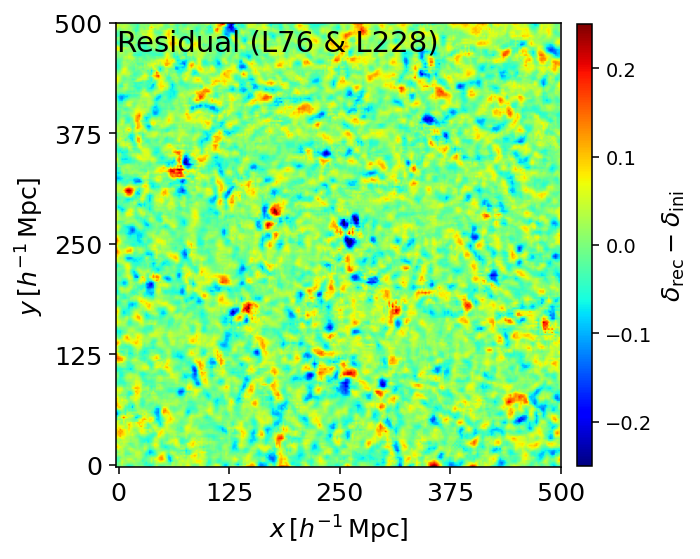} 
\\
 \end{tabular}
 \caption{
 Maps of density fluctuation field in a $1.95\hmpc$ slice are shown for the following cases: \textit{Upper left}: initial condition ($z=10$); \textit{Upper right}: the residual between reconstruction and initial condition $\delta_\mathrm{rec}-\delta_\mathrm{ini}$ using a sub-box of $\lsub\sim76\hmpc$; \textit{Lower left} : the residual using  $\lsub\sim228\hmpc$. 
 \textit{Lower right} : the residual using both $\lsub\sim\{76,228\}\hmpc$. We smoothed both initial and reconstructed density fields with a Gaussian filter with $\sigma=3\hmpc$, as described in Section~\ref{subsec:dataset}.
}
 \label{fig:map}
\end{figure*}

In Fig.~\ref{fig:map}, we show the density maps of an initial condition at $z=10$ and residual maps, defined as $\delta_\mathrm{rec}-\delta_\mathrm{ini}$.
Both the single-input and dual-input models succeed in recovering the large-scale structure of the initial density field within residual $\sim 0.2$. Notably, systematic patterns emerge in the residuals: high-density regions in the initial condition tend to show negative residuals, while low-density regions tend to show positive ones. This indicates that the reconstructions tend to underestimate peaks and overestimate voids.  
Among the models, the dual-input model ($\lsub\sim\{76,228\}\hmpc$) shows the smallest residuals across both high- and low-density regions, suggesting it better captures the dynamic range of the initial density field than single-input models.
This appearance is consistent with PDFs discussed in Section~\ref{subsec:result_optimal_scale}, which show that the single-input model tends to "pull back" density fluctuations excessively, leading to a narrower PDF than the target.


\section{Discussions}
\label{sec:discussions}
Our analysis demonstrates that the scale of the input sub-box plays a crucial role in the performance of CNN-based reconstruction. 
The optimal sub-box size, according to multiple diagnostic metrics, is found to be $\lsub = 150$–$200\ \hmpc$.

This result reflects a fundamental trade-off between resolution and the global structure.
Smaller sub-boxes ($\lsub\lesssim100\hmpc$) provide higher spatial resolution but lack large-scale context, making it harder for the network to distinguish local fluctuations from global structures. On the other hand, 
large sub-boxes ($\lsub\gtrsim250\hmpc$) include broader context but dilute small-scale details, limiting the network's ability to resolve fine features in the initial conditions. The intermediate scale of $\lsub\sim150$–$200\hmpc$ appears to strike the right balance in standard cosmological models: it is large enough to capture meaningful environmental information, yet still preserves the local detail necessary for accurate reconstruction.

On small scales ($0.2<k<0.4,h,\mathrm{Mpc}^{-1}$), we observe that different sub-box sizes yield optimal results depending on the evaluation metric. Specifically, sub-boxes with $\lsub\sim114$–$190\hmpc$ best reproduce the transfer function, indicating accurate recovery of the overall amplitude of the power spectrum. In contrast, slightly larger sub-boxes with $\lsub\sim190$–$226\hmpc$ achieve higher correlation coefficients, suggesting better phase alignment between the reconstructed and true initial fields. 
This difference likely reflects a trade-off between resolution and contextual information: smaller sub-boxes emphasize local structure and detail, improving power recovery, while larger sub-boxes provide a broader context that helps preserve large-scale coherence and phase information. The correlation coefficient is more sensitive to the alignment of features, whereas the transfer function reflects only amplitude matching. Thus, these metrics highlight complementary aspects of reconstruction performance, and the ideal sub-box size may differ depending on the scientific objective.

Interestingly, the best-performing sub-box size is significantly larger than the $\sim76\hmpc$ scale used in earlier studies such as \citet{Mao2021}, highlighting the benefit of incorporating wider spatial context in single-input models. However, it is important to note that the optimal scale identified here likely depends on the specific CNN architecture and training hyper-parameters used in this work. We did not perform an extensive hyperparameter search or architectural tuning; thus, our results should be interpreted as representative rather than definitive. Further improvements in reconstruction quality may be possible with more carefully optimized network designs tailored to different input scales.

The success of our dual-input 3D CNN model highlights the critical role of multi-scale information in the reconstruction of cosmological initial conditions. While previous models based on single-input sub-boxes can achieve a degree of success, they remain fundamentally limited by the trade-off between local resolution and global context. Small sub-boxes provide fine spatial detail but suffer from boundary effects and limited contextual information. Conversely, large sub-boxes capture broad-scale structure but tend to suppress local fluctuations due to spatial averaging. Our dual-input approach effectively overcomes this limitation by combining both types of information, enabling the network to integrate fine and coarse features simultaneously. Additionally, as noted by \citet{Shallue2023}, CNNs are fundamentally local models and not well-suited for capturing long-range interactions such as gravity. In our framework, the lower-resolution input at $\lsub = 228~\hmpc$ is intended to extract information from large-scale modes, while the higher-resolution input at $\lsub = 76~\hmpc$ provides small-scale structure, enabling the model to approximate both local and non-local effects. Conceptually, this approach parallels their hybrid strategy combining CNNs with density reconstruction based on the Zel'dovich approximation \citep{Eisenstein2007}, where large-scale displacements are modeled analytically while local corrections are learned via machine learning.

The improvement in the reconstruction of small-scale structures is particularly noteworthy. The dual-input model maintains both high transfer function and high correlation coefficient at intermediate and small scales ($k \gtrsim 0.2\ihmpc$), where single-input models tend to fail due to the loss of phase coherence and power suppression. This suggests that the model is able to recover not just the amplitude but also the spatial configuration of the underlying modes with higher fidelity. 



Another advantage of the dual-input model is its more efficient use of available data. Although both the single- and dual-input models ultimately draw information from the same parent simulation box ($1 \hgpc$ per side), the dual-input model leverages this information more effectively by combining inputs from different sub-box sizes. This allows the network to capture both large-scale context and small-scale details simultaneously, without requiring additional simulation volume. In this sense, the dual-input model maximizes the information extracted from a fixed dataset, making it a more data-efficient architecture. Furthermore, scaling up to $\lsub = 228~\hmpc$ with a fixed grid resolution of $1.95\hmpc$ in a single-input model would require approximately $3^3 = 27$ times more memory per input sample, significantly increasing the computational cost for CNN training and inference. In contrast, our dual-input model requires only about twice the memory per sample while effectively leveraging multi-scale information. This design aims to efficiently utilize information at multiple scales within limited machine resources.

In summary, the dual-input model offers a simple but powerful improvement by combining local and global information. This multi-scale approach is a natural match for the hierarchical nature of cosmic structure and can be a useful foundation for future work in cosmological reconstruction.

\section{Conclusions}
In this paper, we investigated how the choice of input sub-box size $\lsub$ affects the performance of single-input 3D CNN models in reconstructing the initial density field from the late-time matter distribution. Using a range of sub-box sizes from $38$ to $380\hmpc$, we evaluated reconstruction accuracy through multiple statistical metrics, including the loss function, probability density functions, transfer functions, and correlation coefficients. Our results consistently show that reconstruction performance improves with increasing $\lsub$ up to an intermediate scale ($\lsub\sim152\hmpc$), beyond which it gradually deteriorates. This trend suggests that there exists an optimal scale that balances local resolution and the inclusion of broader environmental context. Interestingly, the optimal $\lsub$ depends slightly on the evaluation metric: for instance, the transfer function peaks at $\lsub\sim114$–$190\hmpc$, while the correlation coefficient is highest at $\lsub\sim190$–$226\hmpc$. These findings highlight the importance of carefully selecting the input scale when applying deep learning to cosmic structure reconstruction. We emphasize that our analysis used a fixed CNN architecture and hyper-parameters; the exact optimal scale may shift with network design or training strategy, and further tuning could yield improved performance.

In the second part, we developed and tested a dual-input 3D CNN model that takes sub-boxes of two different sizes ($\lsub\sim76$ and $228\hmpc$) as inputs, aiming to simultaneously capture small-scale resolution and large-scale context. Compared to single-input models, our dual-input architecture significantly improves reconstruction performance across multiple metrics—including lower loss, better-matched PDFs, higher transfer functions, and improved correlation coefficients, particularly on small scales ($k\gtrsim0.2\ihmpc$). Notably, the dual-input model achieves these improvements while using the same total amount of information from the parent simulation volume, making more efficient use of available data. These results demonstrate that combining information across scales within a single network is a powerful approach for reconstructing the initial density field and opens a promising direction for future work in cosmological inference using deep learning.

Together, these results emphasize that both the choice of input scale and the use of multi-scale architectures are crucial for accurate cosmological field reconstruction. Our findings suggest that future deep-learning approaches can benefit from incorporating multi-scale information to optimally recover the initial conditions of the Universe from its evolved structure. Building on this work, future directions include extending our method to redshift space, where distortions complicate the observed density field, and evaluating its performance in reconstructing halo and galaxy distributions to enable direct applications to observational data. While \citet{Parker2025} focus on observationally motivated reconstruction of halos and galaxies using a hybrid method that combines standard BAO reconstruction with subgrid-based deep learning, our work complements theirs by systematically investigating how architectural design and input resolution impact CNN-based reconstruction from the matter density field. Together, these approaches represent complementary pathways toward achieving accurate and scalable initial condition recovery for next-generation cosmological surveys.

Additionally, integrating physical constraints into the training process and exploring hybrid models that combine deep learning with traditional reconstruction techniques may further enhance interpretability and robustness.

\label{sec:conclisions}

\section*{Acknowledgements}
We thank Yin Li for insightful and helpful discussions that contributed to the development of this work. This work is supported by JSPS Kakenhi Grant Numbers: JP22K21349, JP23H00108, 25H01551 (AJN), 21H04467, and 24K00625 (KI). It is also supported by the JST FOREST Program JPMJFR20352935 and the JSPS Core-to-Core Program (grant numbers: JPJSCCA20200002 and JPJSCCA20200003).


\section*{Data Availability}
The data underlying this article will be shared on reasonable request to the corresponding author.
 



\bibliographystyle{mnras}
\bibliography{references} 




\appendix

\section{Comparison between three pairs of dual-input model}
\label{sec:app_multi}
In the main text, we presented the results of the dual-input model using sub-box sizes of $\lsub \sim \{76, 228\}~\hmpc$, demonstrating its performance in terms of reconstruction quality. In this appendix, we extend the analysis by comparing three possible scale combinations. We aim to better understand the relative strengths of each scale combination and the roles that small- and large-scale features play in the network's ability to recover the initial density field.

\begin{figure}
\centering
    \centering
    \includegraphics[width=0.9\columnwidth]{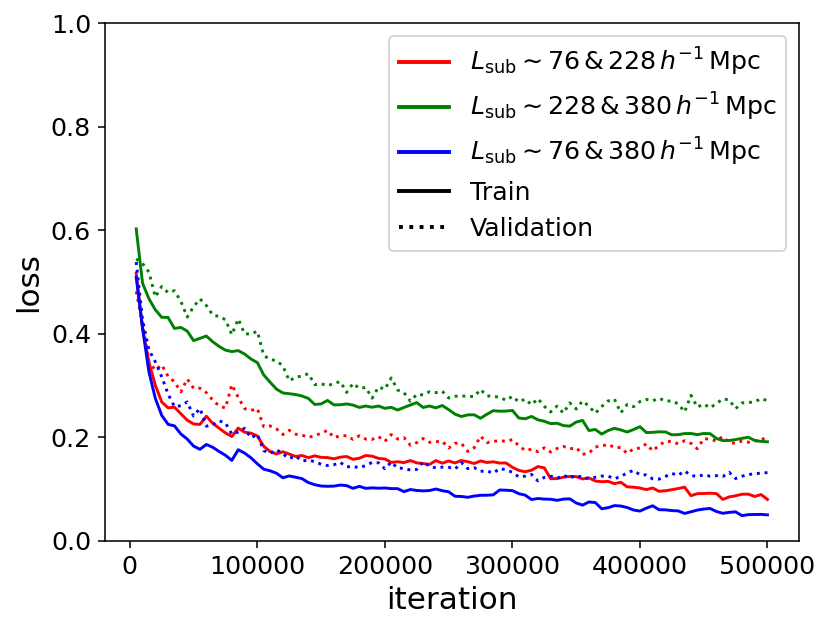}
    \caption{Loss functions as defined in equation~(\ref{eq:loss_function}). Results are shown for three scale pairs of dual-input model: $\lsub~[\hmpc] \sim $ \{76, 228\} (red), \{228, 380\} (green) and \{76, 380\} (blue).}
    \label{fig:loss_add}
\end{figure}

\begin{figure}
    \centering
    \includegraphics[width=0.9\columnwidth]{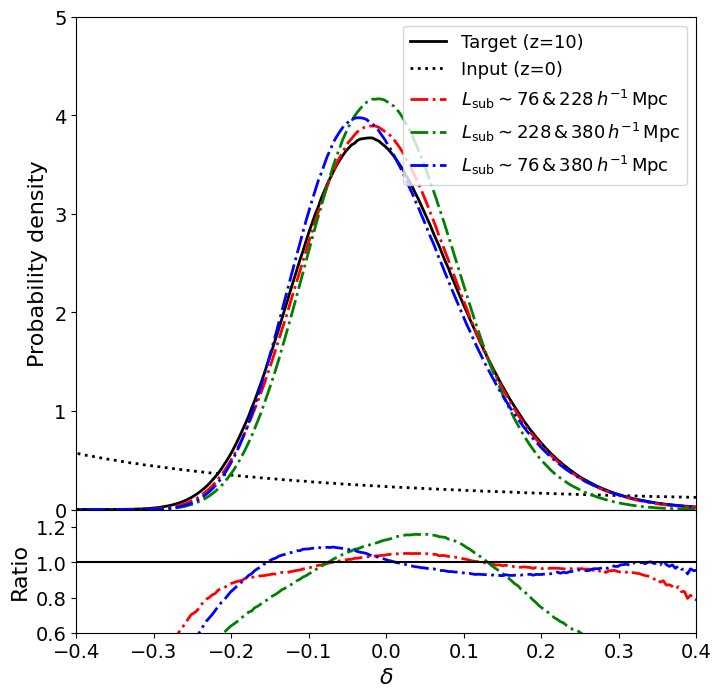}
    \caption{PDFs of the density fluctuation $\delta$ and ratio between a target and three outputs. Results are shown for three scale pairs of dual-input model: $\lsub~[\hmpc] \sim $ \{76, 228\} (red), \{228, 380\} (green) and \{76, 380\} (blue).}
    \label{fig:pdf_add}
\end{figure}

\begin{figure}
\centering
    \centering
    \includegraphics[width=0.9\columnwidth]{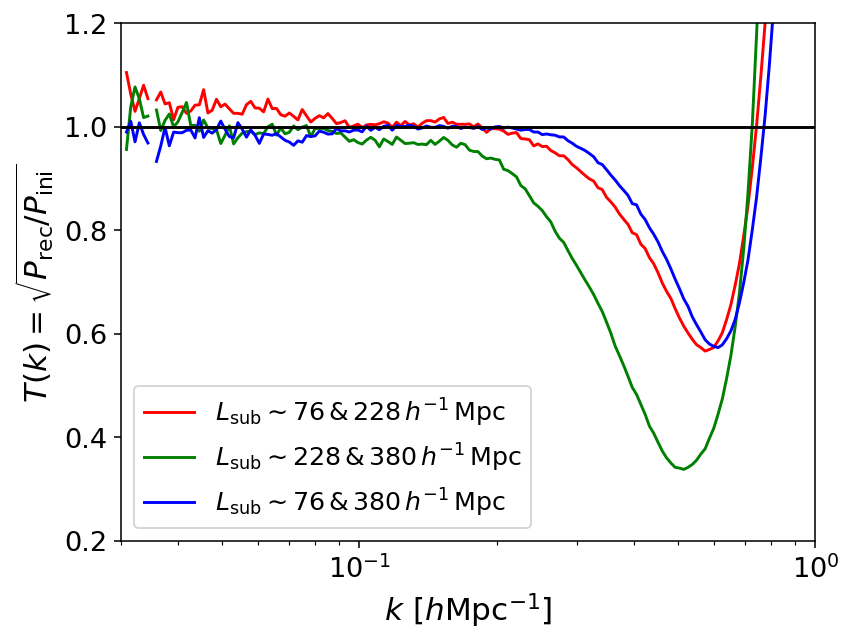}
    \caption{Transfer function between the power spectrum of the reconstructed and target initial density fields, defined in equation~\ref{eq:transfer}. Results are shown for three scale pairs of dual-input model: $\lsub~[\hmpc] \sim $ \{76, 228\} (red), \{228, 380\} (green) and \{76, 380\} (blue).}
    \label{fig:tk_add}
\end{figure}

\begin{figure}
    \centering
    \includegraphics[width=0.9\columnwidth]{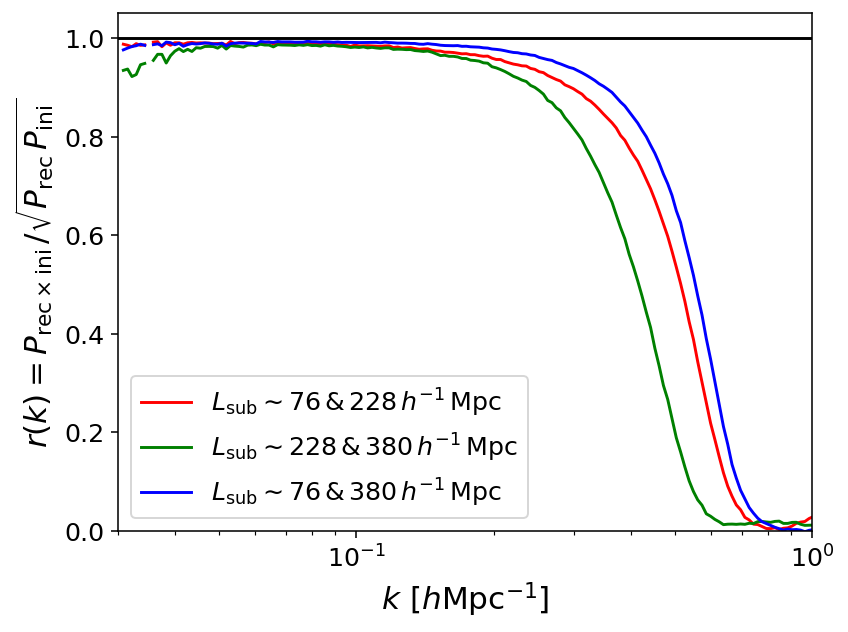}
    \caption{Correlation coefficient between the power spectrum of the reconstructed and target initial density fields, defined in equation~\ref{eq:correlation}. Results are shown for three scale pairs of dual-input model: $\lsub~[\hmpc] \sim $ \{76, 228\} (red), \{228, 380\} (green) and \{76, 380\} (blue).}
    \label{fig:rk_add}
\end{figure}

We evaluate the performance of the dual-input model using three scale pairs: $\lsub~[\hmpc] \sim $ \{76, 228\}, \{228, 380\}, and \{76, 380\} as input. 
Figure~\ref{fig:loss_add} shows the loss function. We compute the mean and standard deviation of the validation loss over iterations $[4,5]\times10^5$, obtaining $0.190 \pm 0.007$, $0.267 \pm 0.008$, and $0.126 \pm 0.004$, respectively. Based on this criterion, the \{76, 380\} pair achieves the best performance.
Figure~\ref{fig:pdf_add} exhibits the PDFs of the density fluctuations. Quantitatively, the KL divergence values for the three scale pairs are 0.0027, 0.0259, and 0.0047, respectively, indicating that the \{76, 228\} combination best matches the true PDF.
These two evaluation criteria, validation loss and KL divergence, are not entirely consistent. 
We also analyse the transfer function and correlation coefficient, shown in Fig.~\ref{fig:tk_add} and Fig.~\ref{fig:rk_add}, both of which indicate that the \{76, 380\} pair achieves the best overall reconstruction quality. This aligns with the trend seen in the validation loss and supports the effectiveness of combining the smallest and largest available scales.




\bsp	
\label{lastpage}
\end{document}